\pdfoutput=1
\documentclass[preprint,12pt]{elsarticle}

\usepackage{graphicx,hyperref}

\usepackage{amssymb}
\usepackage{amsmath}
\usepackage{ifthen}

\usepackage{scalefnt}

\newcommand{\higgsstrahlung}{Higgs Strahlung}
\newcommand{\topdia}{{\abbrev\sf top}}
\newcommand{\dy}{{\abbrev\sf DY}}
\newcommand{\gghz}{{\abbrev\sf ggHZ}}
\newcommand{\gfermi}{G_\text{F}}

\newcommand{\vhnnlo}{{\tt vh@nnlo}}
\newcommand{\zwprod}{{\tt zwprod}}
\newcommand{\code}{\tt}
\newcommand{\abbrev}{\scalefont{.9}}
\newcommand{\cms}{center-of-mass}

\newcommand{\eqn}[1]{Eq.\,(\ref{#1})}
\newcommand{\fig}[1]{Fig.\,\ref{#1}}

\newcommand{\dd}{{\rm d}}
\newcommand{\deriv}[2]{\frac{\dd #1}{\dd #2}}
\newcommand{\order}[1]{{\cal O}(#1)}
\newcommand{\lhc}[1]{{\abbrev LHC#1}}
\newcommand{\qcd}{{\abbrev QCD}}
\newcommand{\sm}{{\abbrev SM}}

\newcommand{\pdf}{{\abbrev PDF}}
\newcommand{\lo}{{\abbrev LO}}
\newcommand{\nlo}{{\abbrev NLO}}
\newcommand{\nnlo}{{\abbrev NNLO}}

\newcommand{\muF}{\mu_\text{F}}
\newcommand{\muR}{\mu_\text{R}}
\newcommand{\mhiggs}{M_\text{H}}
\newcommand{\mv}{M_\text{V}}
\newcommand{\mtop}{M_\text{t}}

\newcounter{bla}

\journal{Computer Physics Communications}

\begin{document}

\begin{frontmatter}



\title{
\vspace*{-6em}
  \begin{flushright}
    {\sf\small 18 Oct 2012 --- WUB/12-20}
  \end{flushright}
  \vspace*{2em} 
{\tt vh@nnlo} --- Higgs Strahlung at hadron colliders\footnote{%
The program is available from\\
\href{http://particle.uni-wuppertal.de/harlander/software/vh@nnlo}{%
\tt http://particle.uni-wuppertal.de/harlander/software/vh@nnlo}. }}

\author[a]{Oliver Brein}
\author[b]{Robert V. Harlander}
\author[b]{Tom J.E. Zirke}

\address[a]{Gro\ss{}karlbacher Stra\ss{}e 10, D-67256 Weisenheim am
  Sand, Germany}
\address[b]{Fachbereich C, Bergische Universit\"at Wuppertal, 42097
  Wuppertal, Germany}

\begin{abstract}
A numerical program for the evaluation of the inclusive cross section
for associated Higgs production with a massive weak gauge boson at
hadron colliders is described, $\sigma(pp/p\bar p\to HV)$,
$V\in\{W,Z\}$. The calculation is performed in the framework of the
Standard Model and includes next-to-next-to-leading order \qcd{} as well
as next-to-leading order electro-weak effects.
\end{abstract}


\end{frontmatter}





{\bf PROGRAM SUMMARY}

\begin{small}
\noindent
{\em Program Title:} {\tt vh@nnlo} \\
{\em Distribution format:} {\tt tar.gz}           \\
{\em Programming language:} {\tt Fortran 77, C++.}           \\
{\em Computer:} {\tt Personal computer.}                          \\
{\em Operating system:} {\tt Unix/Linux, Mac OS.}                                    \\
{\em RAM:} {\tt A few 100 MB.}                                              \\
{\em External routines/libraries:} {\tt LHAPDF (http://lhapdf.hepforge.org/), \\
     CUBA (http://www.feynarts.de/cuba/)}               \\
{\em Nature of problem:}\\
Calculation of the inclusive total cross section for associated Higgs-
and $W$- or $Z$-boson production at hadron colliders through
next-to-next-to-leading order \qcd{}.
   \\
{\em Solution method:}\\
Numerical Monte Carlo integration.
   \\
{\em Running time:}\\
A few seconds for a single set of parameters.
   \\

\end{small}




\section{Introduction}
\begin{figure}
\begin{center}
\begin{tabular}{cc}
\includegraphics[width=0.4\textwidth,bb=100 540 510
  780]{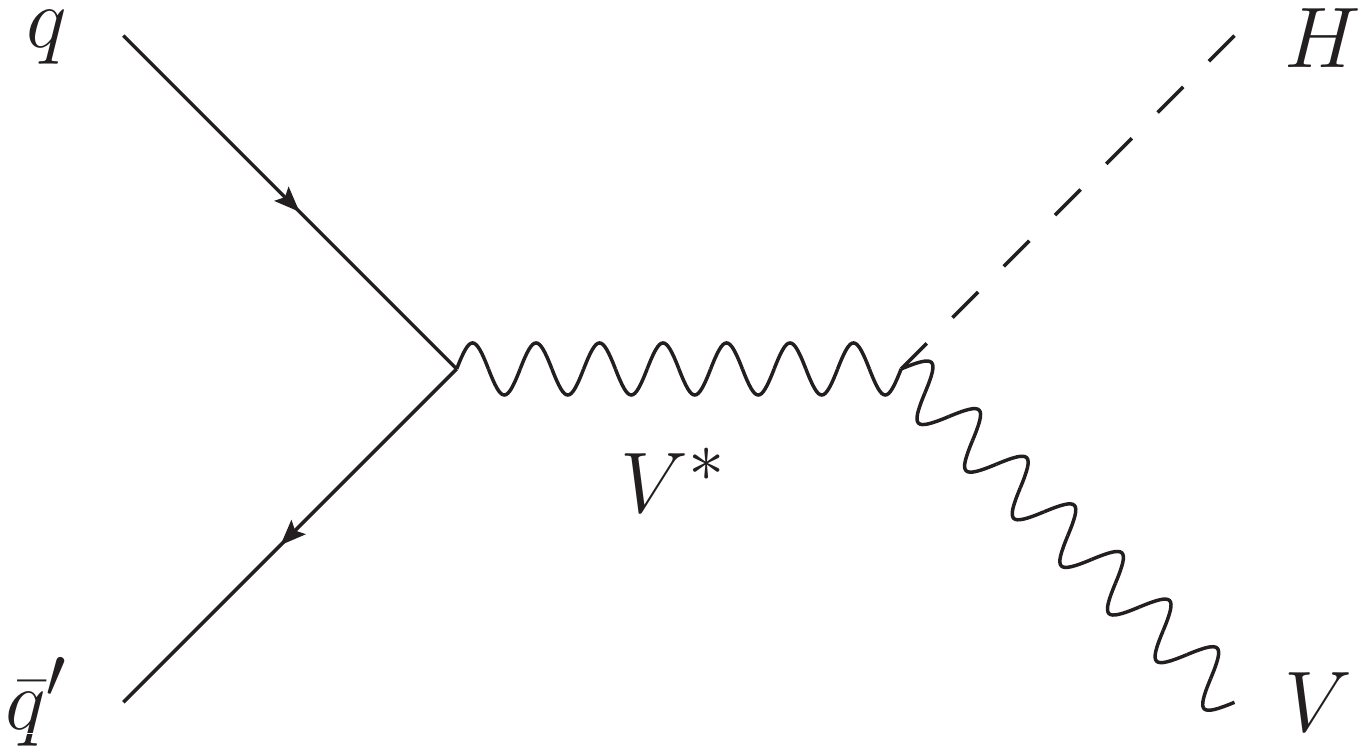} &
\includegraphics[width=0.4\textwidth,bb=100 540 510
  780]{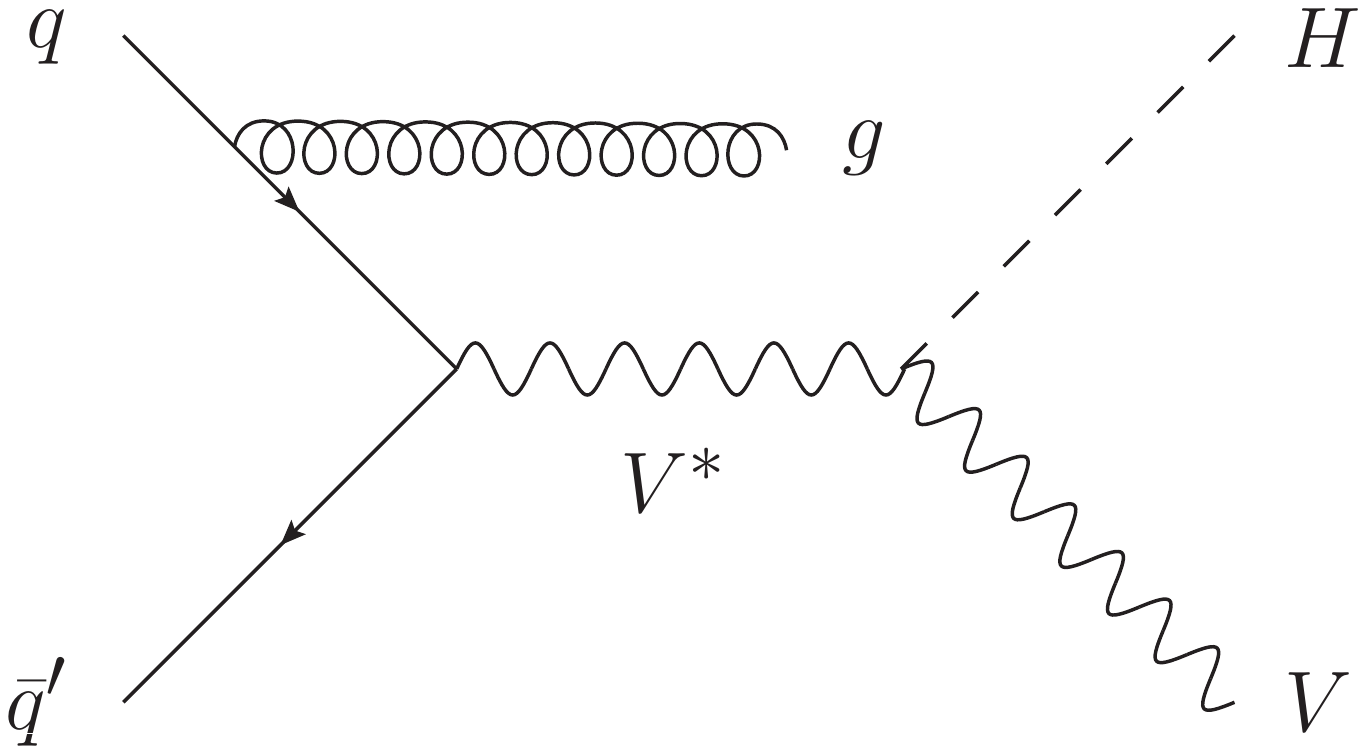}\\
(a) & (b)
\end{tabular}
\caption{Tree-level Feynman diagrams contributing to the processes
  (a) $q\bar q\to V\!H$ and (b) $q\bar q\to V\!Hg$. By crossing, one also
  obtains the diagrams contributing to $qg\to V\!Hq$ and $\bar qg\to
  V\!H\bar q$.}
\label{fig::tree}
\end{center}
\end{figure}

The associated production of a Higgs ($H$) and a weak gauge boson ($V\in
\{W,Z\}$), or ``\higgsstrahlung{}'' for short, has already proven to be
an important process at hadron colliders (see, e.g.,
Refs.\,\cite{Aad:2012gk,Chatrchyan:2012gu,Aaltonen:2012qt}). In the
Standard Model (\sm{}), the leading order (\lo) amplitude is given by
the Feynman diagram shown in \fig{fig::tree}\,(a). We count it as order
$g^4$, where $g$ is the weak gauge coupling.  Through
next-to-leading order (\nlo) \qcd{}, $\order{g^4\alpha_s}$, it can be
represented as the convolution of the production of a virtual gauge
boson $V^\ast$ with the decay rate of $V^\ast$ into the on-shell gauge
boson $V$ and the Higgs boson $H$, schematically: $\sigma_{V\!H} =
\sigma_{V^\ast}\otimes \Gamma_{V^\ast\to V\!H}$. The precise formula
will be given below.

The bulk of the next-to-next-to-leading order (\nnlo) \qcd{}
corrections, $\order{g^4\alpha_s^2}$, is given by the same formula and
will be referred to as \dy{} (for ``Drell-Yan like'') terms in the
following.

\begin{figure}
\begin{center}
\begin{tabular}{ccc}
\includegraphics[width=0.26\textwidth,bb=140 540 460
  780]{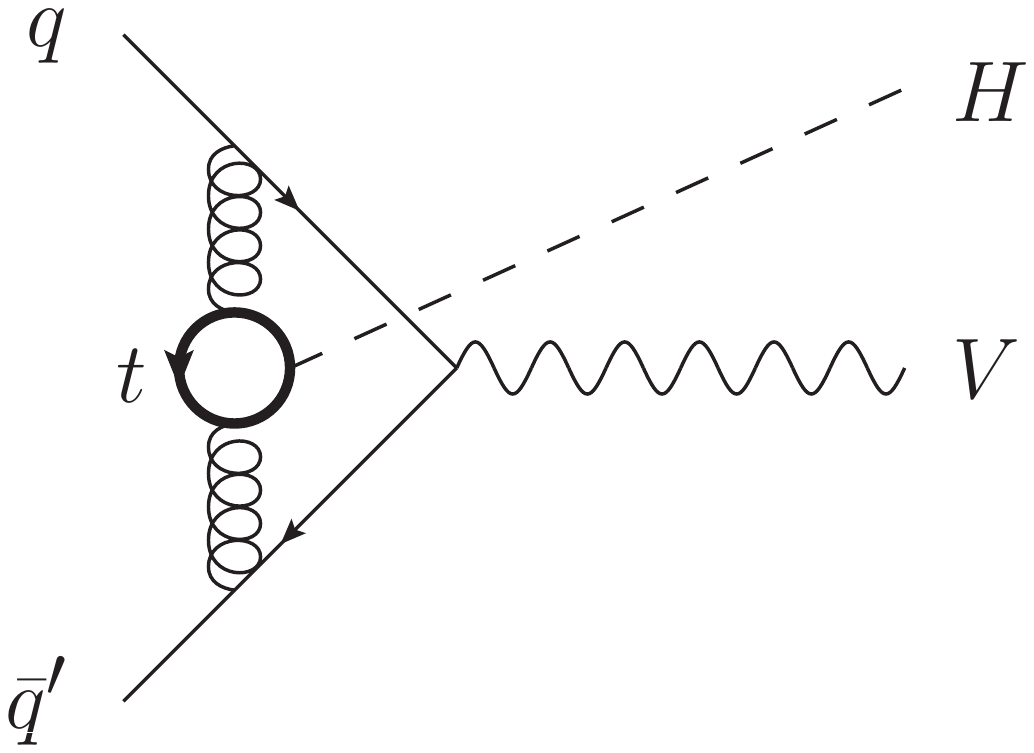} &
\hspace{0.5cm} \includegraphics[width=0.26\textwidth,bb=140 540 460
  780]{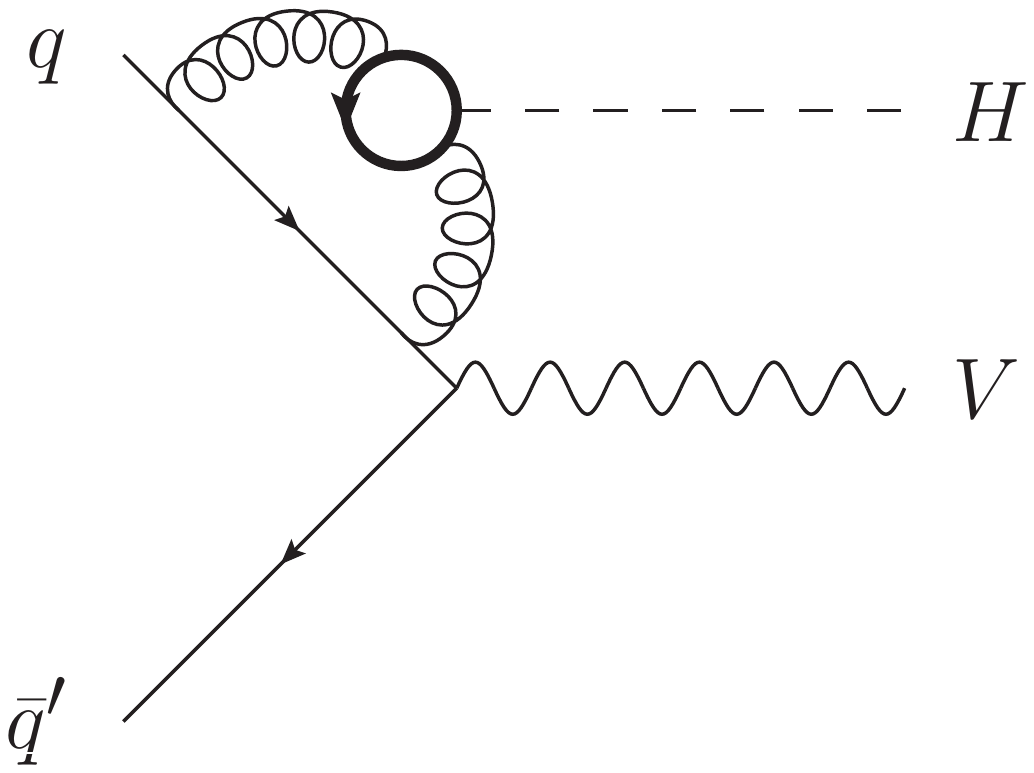} &
\hspace{0.5cm} \includegraphics[width=0.26\textwidth,bb=140 540 460
  780]{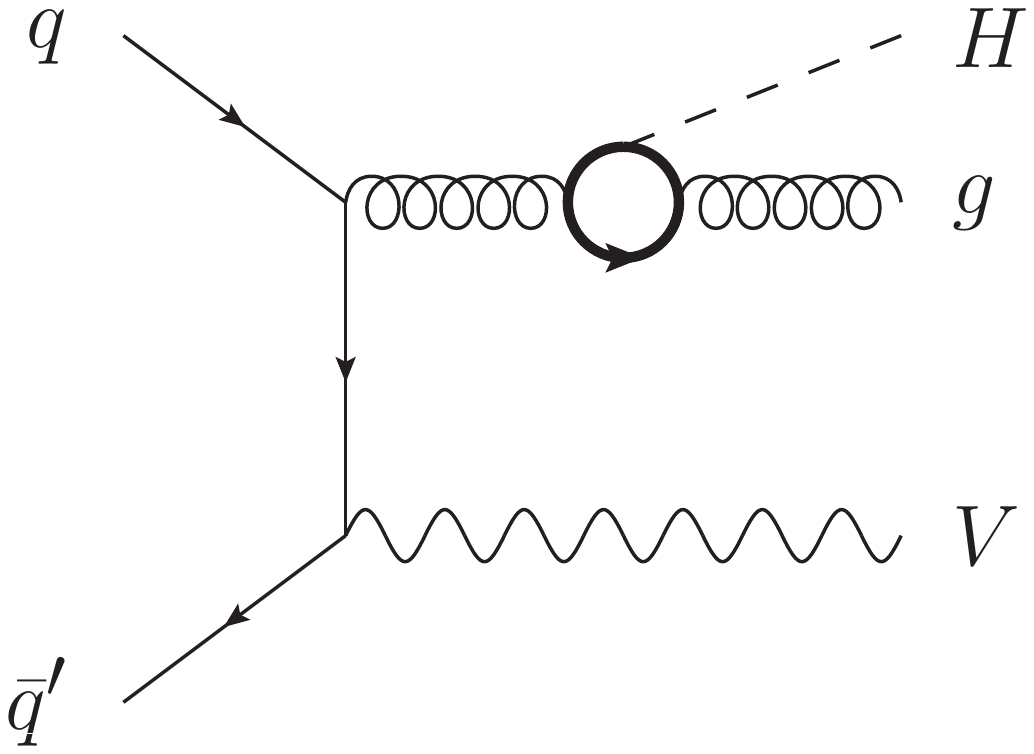} \\
(a) & (b) & (c)
\end{tabular}
\caption{Sample diagrams giving rise to the
  $\order{g^3\lambda_t\alpha_s^2}$ terms in the cross section}
\label{fig::ltas2}
\end{center}
\end{figure}

\begin{figure}
\begin{center}
\begin{tabular}{ccc}
\includegraphics[width=0.26\textwidth,bb=140 540 460
  780]{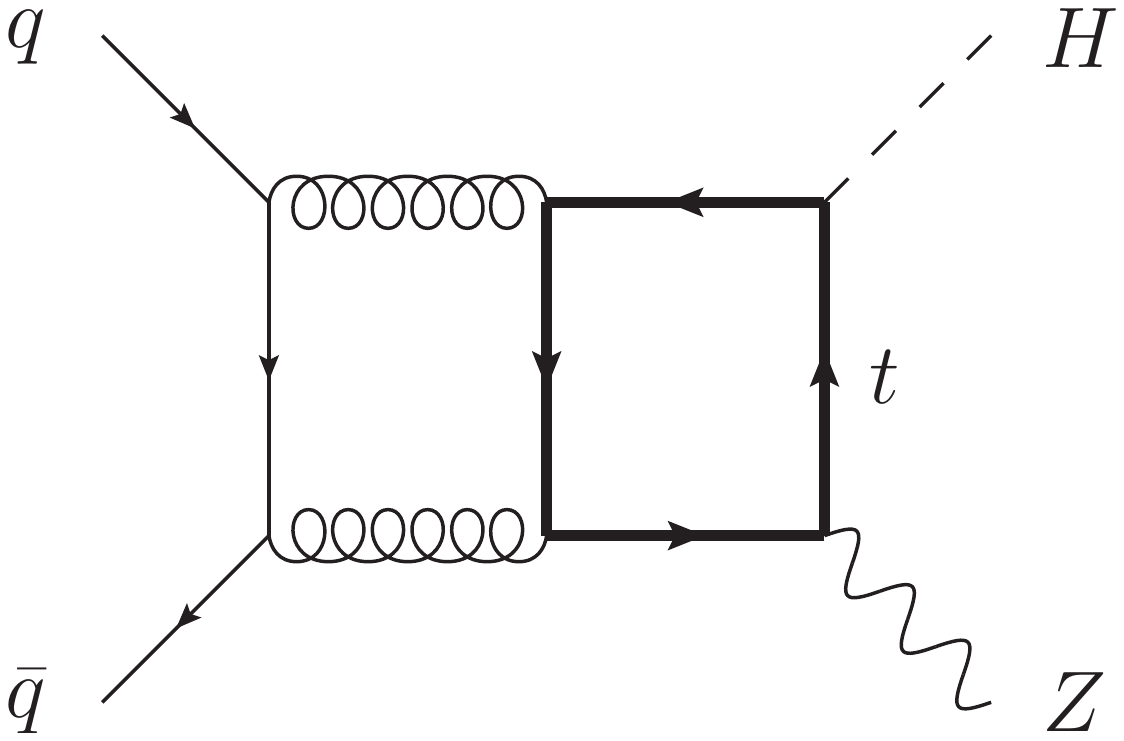} &
\includegraphics[width=0.26\textwidth,bb=140 540 460
  780]{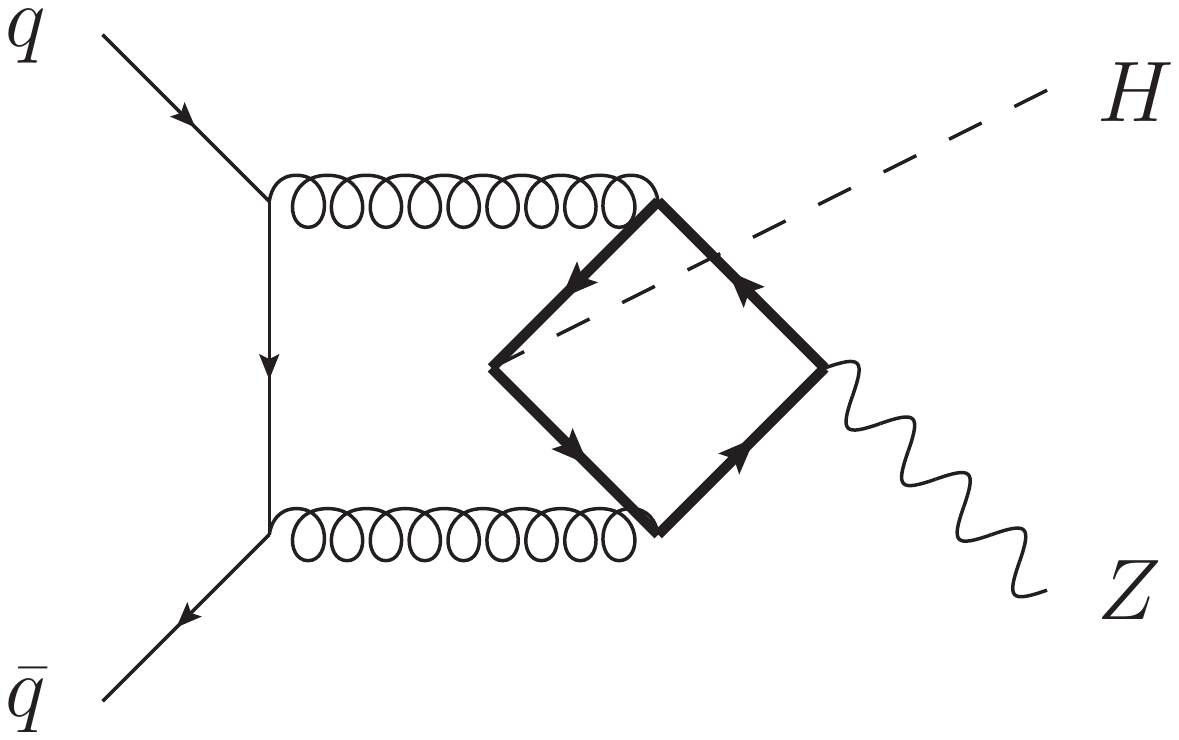} &
\hspace{0.5cm} \includegraphics[width=0.26\textwidth,bb=140 540 460
  780]{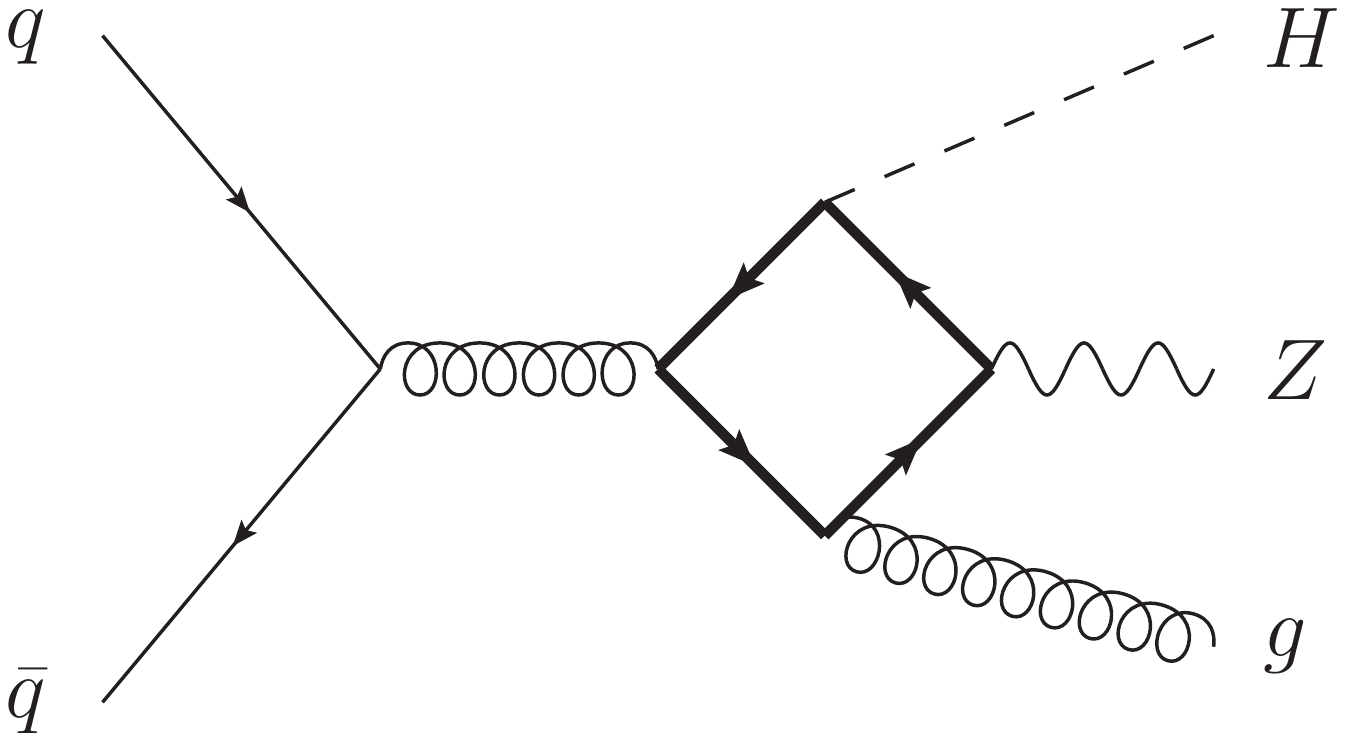} \\
(a) & (b) & (c)
\end{tabular}
\caption{A $\order{g^3\lambda_t\alpha_s^2}$ contributions only present for
  $Z\!H$ production. Diagram (c) gives rise to the $q\bar q\to
  Z\!Hg$, $qg\to Z\!Hq$, and $\bar qg\to Z\!H\bar q$ processes.}
\label{fig::zhaa}
\end{center}

\end{figure}
\begin{figure}
\begin{center}
\begin{tabular}{c}
\includegraphics[width=0.4\textwidth,bb=100 540 510
  780]{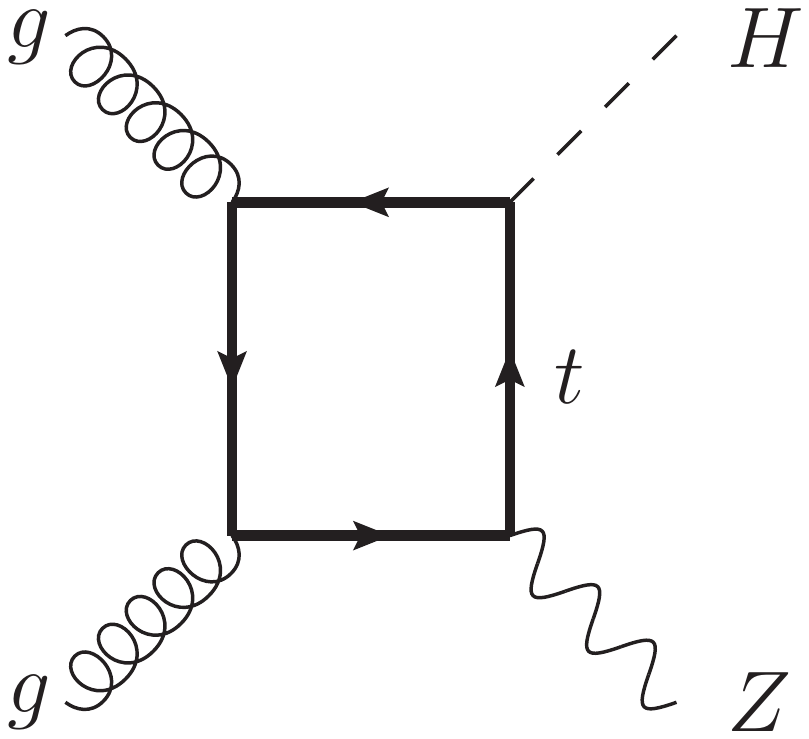}
\end{tabular}
\caption{Typical diagram for the \gghz{} contribution.}
\label{fig::gghz}
\end{center}
\end{figure}

At $\order{\alpha_s^2}$, there are contributions to the cross section
which cannot be derived from the Drell-Yan process as before. One such
class is given by diagrams where the Higgs boson is radiated off a
virtual top-quark loop which is inserted either in a real or a virtual
gluon line. Sample diagrams are shown in \fig{fig::ltas2}.  Apart from
that, for $Z\!H$ production, both the $Z$ and the $H$ can couple to the
top-quark loop; examples are shown in \fig{fig::zhaa}.  All these
diagrams enter the total cross section at
$\order{g^3\lambda_t\alpha_s^2}$ through the interference with the \lo{}
amplitudes shown in \fig{fig::tree}. Although these are not the only
contributions induced by top-quark loops, we will refer specifically to
these $\order{g^3\lambda_t\alpha_s^2}$ terms as the ``\topdia{}'' terms.

For $Z\!H$ production, yet another top-quark induced contribution enters
at $\order{\alpha_s^2}$. Sample diagrams are shown in
\fig{fig::gghz}. They differ from the previous ones in that they contain
two initial state gluons. Since this does not occur at lower orders in
$\alpha_s$, it is the square of the corresponding amplitude that enters
the cross section, rather than the interference with lower order terms.
It will be referred to as the ``\gghz{}'' contribution in what
follows. Note that there are also \dy{} diagrams with $gg$ initial state,
but they belong to the double real emission terms (e.g.\ $gg\to
V\!Hq\bar q$), while the \gghz{} contribution is a purely virtual
correction. In our counting of coupling constants, it enters the cross
section at $\order{g^2\lambda_t^2\alpha_s^2}$.

The \nlo{} \qcd{} corrections have been known for a long
time~\cite{Han:1991ia}. The \dy{} \nnlo{} corrections, based on the
\nnlo{} results for the actual Drell-Yan process~\cite{Hamberg:1991np,Harlander:2002wh}
and including the \gghz{} terms~\cite{Kniehl:1990iv}, were first
presented in Ref.\,\cite{Brein:2003wg}. The \topdia{} effects were
subsequently evaluated in Ref.\,\cite{Brein:2011vx}. For a Higgs mass of
$\mhiggs=125$\,GeV, inclusion of the \nlo{} corrections enhances the
total inclusive cross section for $W\!H$ production by about 31\% (27\%,
41\%) at the \lhc{8} (\lhc{14}, Tevatron); the effect of the \nnlo{}
\dy{} terms is an additional 3\% (3\%, 10\%), and the \topdia{} effects
amount to about 1\% in all three cases.  For $Z\!H$
production, these numbers are very similar; the additional \gghz{}
component amounts to about 5\% (9\%, 0.2\%).

These numbers show that the numerical effect of the \nnlo{} \dy{}
corrections is rather small; however, including them significantly
reduces the theoretical uncertainty due to scale
variations\,\cite{Brein:2003wg}, in particular for the $W\!H$
process. For $Z\!H$ production, on the other hand, the fact that the
\gghz{} terms are separately finite, gauge independent, and proportional
to $\alpha_s^2(\mu)$ introduces an additional and significant amount of
theory uncertainty (see also
Refs.\,\cite{Dittmaier:2012vm,Dittmaier:2011ti}).

Currently, the program \vhnnlo{} includes the \dy{}, the \topdia{}, and
the \gghz{} contributions at $\order{\alpha_s^2}$.  Electro-weak
corrections to the Higgs Strahlung process are of the order of 5\% and
thus larger than the \qcd{} scale
uncertainty~\cite{Ciccolini:2003jy,Brein:2004ue}. Since they do not
depend on any of the \qcd{} parameters like \pdf{}s or the strong
coupling, we included them in {\tt vh@nnlo} in the form of a correction
factor by linearly interpolating the numbers given in Ref.\cite{lhxswg}.

Note that {\tt vh@nnlo} only evaluates the total inclusive cross
section; kinematical distributions, for example in the Higgs' transverse
momentum or rapidity cannot be calculated with this program. We refer
the reader to Refs.\,\cite{Ferrera:2011bk,Denner:2011id,Dawson:2012gs} (and
references therein) for that purpose.


\section{Description of the program}


\subsection{Structure}
This section describes the internal structure of \vhnnlo{}.


\subsubsection{\dy{} terms}
The core of the program consists of the {\code Fortran 77} code
\zwprod{} by W.~van Neerven which is publicly available from
Ref.\,\cite{zwprod}. It evaluates the production of an off-shell weak
gauge boson $V^\ast$ of invariant mass $q^2$ through \nnlo{} \qcd{} by
implementing the results for the partonic cross section
$\hat\sigma^{V^\ast}_{\alpha\beta}$ for the process $\alpha\beta\to
V^\ast+X$ ($\alpha,\beta\in\{q,\bar q,g\}$) of
Refs.\,\cite{Hamberg:1991np,Harlander:2002wh} and integrating them over
the parton density functions $\phi_{\alpha/p}$:
\begin{equation}
\begin{split}
  \sigma^{V^\ast}_{pp'}(q^2,s) &\equiv
  \frac{1}{s}\sum_{\alpha,\beta\in\{q,\bar q,g\}}\int_{q^2}^s\dd \hat s
  \, \hat\sigma^{V^\ast}_{\alpha\beta}(q^2,\hat
  s,\muF)\,\\&\hspace*{1em}\times \int_{\hat s}^s\frac{\dd \tilde
    s}{\tilde s}\left[\frac{ \phi_{\alpha/p}(\tilde s/s,\muF)
      \phi_{\beta/p'}(\hat s/\tilde
      s,\muF)}{1+\delta_{\alpha\beta}}\right]\, \,, \qquad p'\in
  \{p,\bar p\}\,.
\label{eq::sigv}
\end{split}
\end{equation}
The \dy{} terms of the inclusive total cross section for $V\!H$
production are then obtained by folding this expression with the decay
rate of the off-shell vector boson $\dd\Gamma$:
\begin{equation}
\begin{split}
\sigma^\dy_{pp'}(s,\mhiggs^2,\mv^2) = \int_{(\mhiggs+\mv)^2}^s\dd q^2
\sigma^{V^\ast}_{pp'}(q^2,s)\deriv{\Gamma(\mhiggs^2,\mv^2;q^2)}{q^2}\,,
\label{eq::sighv}
\end{split}
\end{equation}
where
\begin{equation}
\begin{split}
\deriv{\Gamma(\mhiggs^2,\mv^2;q^2)}{q^2}
= \frac{\gfermi\mv^4}{2\sqrt{2}\pi^2}
\frac{\lambda^{1/2}(\mv^2,\mhiggs^2;q^2)}{(q^2-\mv^2)^2}\left(
1 + \frac{\lambda(\mv^2,\mhiggs^2;q^2)}{\mv^2/q^2}\right)\,,
\label{eq::gamma}
\end{split}
\end{equation}
and
\begin{equation}
\begin{split}
\lambda(x,y,z) \equiv \left(1-\frac{x}{z}-\frac{y}{z}\right)^2-4\frac{xy}{z}\,.
\end{split}
\end{equation}

In \vhnnlo{}, the program \zwprod{} is modified in such a way that it
performs the three integrations of \eqn{eq::sigv} and (\ref{eq::sighv})
simultaneously using the {\tt Vegas} algorithm~\cite{Lepage:1977sw} as
implemented in the {\tt CUBA} library~\cite{Hahn:2004fe}. 

For the \pdf{}s, \vhnnlo{} links the {\tt LHAPDF} library~\cite{lhapdf}
which allows one to switch between different \pdf{} sets conveniently.
Neither {\tt LHAPDF} nor {\tt CUBA} are part of the distribution of {\tt
  vh@nnlo}; they have to be installed separately.


\subsubsection{\gghz{} terms}
The \gghz{} terms of order $\order{g^2\lambda_t^2\alpha_s^2}$ are
calculated by a routine that was generated with the help of {\tt
  FeynArts}~\cite{Hahn:2000kx}. Due to the top-quark loop, one arrives
at massive box integrals that are evaluated using a slightly modified
version of the {\tt LoopTools/FF}
library~\cite{Hahn:1998yk,vanOldenborgh:1989wn} which is included in the
distribution of \vhnnlo{}. For the \pdf{}s, again {\tt
  LHAPDF}~\cite{lhapdf} is used.


\subsubsection{\topdia{} terms}
The \topdia{} contributions of order $\order{g^3\lambda_t\alpha_s^2}$
are evaluated by a {\code C++} routine. The calculation is based on the
heavy-top limit which was shown to work well within the overall
available accuracy~\cite{Brein:2011vx}.  Folding the partonic
differential cross section with the \pdf{}s and integrating it over the
phase space requires a three (six) dimensional numerical integration for
the virtual (real) contribution, which is again performed using {\tt
  LHAPDF} and the {\tt CUBA} implementation of {\tt VEGAS}.


\subsubsection{Electro-weak corrections}
As already mentioned above, electro-weak corrections are implemented in
the form of a two-dimensional grid spanning over the values
$\mhiggs\in[80,200]$\,GeV in steps of 5\,GeV, $\mhiggs\in[200,300]$\,GeV
in steps of 10\,GeV, and $\sqrt{s} \in\{7,8,9,10,14\}$\,TeV. The
correction factor is very flat; at intermediate values it is therefore
obtained by linear interpolation.


\subsubsection{Total cross section}
In summary, the individual orders of the inclusive Higgs Strahlung cross
section are calculated as follows:
\begin{equation}
\begin{split}
\sigma_\text{\lo} &= (1+\delta_\text{EW})\sigma^\dy_\text{\lo}\,,\\
\sigma_\text{\nlo} &= (1+\delta_\text{EW})\sigma^\dy_\text{\nlo}\,,\\
\sigma_\text{\nnlo} &= (1+\delta_\text{EW})\sigma^\dy_\text{\nnlo}
+ \sigma^\gghz{} + \sigma^\topdia{}\,,\\
\end{split}
\end{equation}
where $\delta_\text{EW}$ denotes the electro-weak correction factor.



\subsection{Installation and compilation}


\subsubsection{Installation}
Unpacking the tar ball will create the following directory
  structure:

{\footnotesize
\begin{verbatim}
vh@nnlo/
   infiles/
   mainfiles/
   manual/
   output/
   source/
   x/
\end{verbatim}}

\begin{description}
\item[\tt vh@nnlo:] this will also be referred to as the root directory
  of \vhnnlo{}. It contains the file \verb/README/ which briefly
  describes the most important topics for installation, compilation, and
  operation of the program \vhnnlo{}.
\item[\tt source:] contains both the sources of \vhnnlo{}, as
well as the main \verb/Makefile/ and a very simple compilation script
named \verb/vh@nnlo_install/.
\item[\tt x:] should be empty upon first installation. It will contain
  the executable once \vhnnlo{} is compiled.
\item[\tt output:] a directory to store output files.
\item[\tt infiles:] a directory to store input files.
\item[\tt manual:] contains this manual.
\item[\tt mainfiles:] a directory to store the main program. Currently,
  it contains only the file {\tt main.f}.
\end{description}


\subsubsection{Compilation}
Before compiling, it might be necessary to adjust some of the library
paths in \verb$source/Makefile$.

Full compilation is most easily done by saying
\begin{verbatim}
./vh@nnlo_install
\end{verbatim}
in the directory \verb/source/. This will compile the
\verb$LoopTools/FF$ library, and the \gghz{}, \dy{}, and \topdia{}
components of the calculation. The main executable \verb/x.main/ will be
copied to the directory \verb/x/ and should be called from the root
directory of \vhnnlo{}.

Sometimes it is obvious that the \gghz{} component does not contribute
to the cross section. This is true for $W\!H$ production, for example,
or if only the \nlo{} result is needed. In that case, one may work with
only a subset of the code and exclude, for example, the
\verb$LoopTools/FF$ library from the compilation. Instead of running the
installation script, one may then simply say
\begin{verbatim}
make GGHZ=no
\end{verbatim}
in the \verb/source/ directory.



\subsection{Operation}


\subsubsection{Input}\label{sec:input}
\vhnnlo{} is controlled by a single input file. An example named {\tt
  in.init} is included in the distribution in the directory {\tt
  infiles}. It is structured like an {\abbrev SLHA} input
file~\cite{Skands:2003cj}, but is not {\abbrev SLHA} compatible (it uses
its own {\tt Blocks}, for example).  The content of {\tt in.init} as
contained in the distribution is:

\newpage
{\footnotesize
\begin{verbatim}
Block REAL
     1  8.000000D+03 # sqrt(s) [GeV]
     2  125.0000D+00 # mh [GeV]
     3  0.100000D+01 # muR/q
     4  0.100000D+01 # muF/q
     8  172.5d0      # mt [GeV]
     9  4.75d0        # mb [GeV]
    11  0.911876D+02 # Mz [GeV]
    12  0.803980D+02 # Mw [GeV]
    13  0.249520D+01 # GammaZ [GeV]
    14  0.214100D+01 # GammaW [GeV]
    15  0.116637D-04 # GFermi [1/GeV^2]
    18  0.508000D-01 # sin^2(thetaC)
Block INTEGER
     1      1 # [1:pp][2:ppbar]
     2      2 # [0:LO][1:NLO][2:NNLO] (adjust PDF set as well!)
     3      1 # [0:WH][1:ZH]
     4      0 # PDF number
     5      5 # nf
     6      1 # [0:w/o][1:with] electro-weak corrections
Block CHAR
     1  MSTW2008nnlo68cl.LHgrid # PDF set
\end{verbatim}
} 

It is divided into ``{\tt Blocks}'' that are characterized by the type of
the input variables they contain (\verb/REAL*8/, \verb/INTEGER/, and
\verb/CHARACTER/). Each line inside a {\tt Block} consists of (from left
to right)
\begin{itemize}
\item at least one whitespace character
\item a label of type {\tt INTEGER}
\item the value of the input variable
\item (optionally) a comment, introduced by a hash symbol (\verb/#/)
\end{itemize}
Quite generally, \vhnnlo{} will ignore any text to the right of the hash
symbol~\verb/#/.

In the following, the individual input parameters are described in more
detail. All masses, energies, and decay widths are to be given in GeV.

\newpage
\begin{description}
\item[\tt Block REAL] --- {\it input variables of type} {\tt REAL*8}\\
\begin{tabular}{rcl}
{\tt 1 } &---& {\it $\sqrt{s}$, the hadronic \cms{} energy}\\
{\tt 2 } &---& {\it $\mhiggs$, the mass of the Higgs boson}\\
{\tt 3 } &---& 
\begin{minipage}[t]{.8\textwidth}
{\it $\muR/\sqrt{q^2}$, the renormalization scale
  relative to $\sqrt{(p_\text{V}+p_\text{H})^2}$}\\
$p_\text{V}$ and $p_\text{H}$ are the 4-momenta of the final state weak
gauge boson and the Higgs, respectively. At \lo{},
$q^2=(p_\text{V}+p_\text{H})^2$ is thus the square of the partonic
center-of-mass energy. 
\end{minipage}
\\
{\tt 4 } &---& 
\begin{minipage}[t]{.8\textwidth}
{\it $\muF/\sqrt{q^2}$, the factorization scale
  relative to $\sqrt{(p_\text{V}+p_\text{H})^2}$}\\
See entry {\tt 3}.
\end{minipage}
\\
{\tt 8 } &---& {\it $\mtop$, the on-shell top quark mass}\\
{\tt 9 } &---& {\it $M_\text{b}$, the on-shell bottom quark mass}\\
{\tt 11} &---& {\it $M_\text{Z}$, the $Z$ boson mass}\\
{\tt 12} &---& {\it $M_\text{W}$, the $W$ boson mass}\\
{\tt 13} &---& {\it $\Gamma_\text{Z}$, the $Z$ boson decay width}\\
{\tt 14} &---& {\it $\Gamma_\text{W}$, the $W$ boson decay width}\\
{\tt 15} &---& 
\begin{minipage}[t]{.8\textwidth}
{\it $\gfermi$, the Fermi constant (in GeV$^{-2}$)}\\ Note that the
electro-weak couplings $g,g'$ are expressed in terms of $\gfermi$,
$M_\text{W}$, and $M_\text{Z}$; in particular, we use $\gfermi =
\pi\alpha_\text{QED}/(\sqrt{2}M_\text{W}^2\sin^2\theta_W)$ and
$\sin^2\theta_W = 1-M_\text{W}^2/M_\text{Z}^2$.
\end{minipage}
\\
{\tt 18} &---& 
\begin{minipage}[t]{.8\textwidth}
{\it $\sin^2\theta_C$, the Cabbibo angle}\\
Mixing of the first and second quark generation with the third
  generation is neglected, i.e.\ {\tt vh@nnlo} assumes $V_{qt} =
  \delta_{qb}$, $q\in\{d,s,b\}$. The error introduced by this
  approximation is completely negligible.
\end{minipage}
\end{tabular}

\newpage
\item[\tt Block INTEGER] --- {\it input variables of type {\tt INTEGER}}\\
\begin{tabular}{rcl}
{\tt 1 } &---& {\it collider type: {\tt 1} $\hat{=}$ $pp$, 
{\tt 2} $\hat{=}$ $p\bar p$}\\
{\tt 2 } &---&
\begin{minipage}[t]{.8\textwidth}
{\it order of calculation: {\tt 0} $\hat{=}$ \lo{}, {\tt 1} $\hat{=}$
  \nlo{}, {\tt 2} $\hat{=}$ \nnlo{}}\\ This parameter affects only the
order of $\alpha_s$ that is taken into account for the partonic cross
section. It has no influence on the order at which $\alpha_s$ or the
\pdf{}s are {\it evolved}. This is determined by the choice of the
\pdf{} set ({\tt Block CHAR, entry 1}).\\[-.8em]
\end{minipage}
\\
{\tt 3} &---& {\it type of process}:
{\tt 0} $\hat{=}$ $W\!H$,
{\tt 1} $\hat{=}$ $Z\!H$\\
{\tt 4} &---& 
\begin{minipage}[t]{.8\textwidth}
{\it number of the \pdf{} set
 member}\\ 
Modern \pdf{} sets usually provide a means to evaluate the
 corresponding uncertainties. For this, they contain several ``member''
 \pdf{}s within one set. The set itself is specified in {\tt Block CHAR,
 entry 1}, see below.\\[-.8em]
  \end{minipage}
\\ {\tt 5} &---& {\it number of flavors taking into account in the
  running of $\alpha_s$}\\
{\tt 6} &---& {\it include electro-weak correction factor: {\tt 0}
  $\hat{=}$ no, {\tt 1} $\hat{=}$ yes}
\end{tabular}
\item[\tt Block CHAR] --- {\it input variables of type {\tt CHARACTER}}\\
\begin{tabular}{rcl}
{\tt 1} &---& \begin{minipage}[t]{.8\textwidth}
{\it the name of the \pdf{} set as defined in {\tt LHAPDF}}\\
The perturbative order of the \pdf{} set will determine the order of
the {\abbrev DGLAP} evolution of the \pdf{}s, and also the order to
which $\alpha_s$ is evolved from $M_\text{Z}$ to $\muR$.
\end{minipage}
\end{tabular}
\end{description}

In order to run the program with this input file, say
\begin{verbatim}
x/x.main infiles/in.init
\end{verbatim}
in the main directory of {\tt vh@nnlo}.


\subsubsection{Output}\label{sec:output}
The input file of section\,\ref{sec:input} will lead to the following
output, written to the file {\tt output/out.vh}:\\[-2em]

{\footnotesize
\begin{verbatim}
# -----------------------
#  +++  vh@nnlo version 1.20  +++  
# Authors: T. Zirke and R.V. Harlander (Drell-Yan part)
#          O. Brein (ggHZ part)
# Based on zwprod.f by W. van Neerven and the paper
#    O. Brein, A. Djouadi, R.V. Harlander,
#    Phys.Lett. B579 (2004) 149, hep-ph/0307206.
# Uses LoopTools by T. Hahn and FF by J. van Oldenborgh.
# -----------------------
# In addition, please cite the following papers:
# Han:1991ia          
# Hamberg:1991np      
# Harlander:2002wh    
# Brein:2003wg        
# Ciccolini:2003jy    
# Kniehl:1990iv       
# -----------------------
Block REAL
     1  0.800000D+04 # sqrt(s) [GeV]
     2  0.125000D+03 # mh [GeV]
     3  0.100000D+01 # muR/q
     4  0.100000D+01 # muF/q
     8  0.172500D+03 # Mt [GeV]
     9  0.475000D+01 # Mb [GeV]
    11  0.911876D+02 # Mz [GeV]
    12  0.803980D+02 # Mw [GeV]
    13  0.249520D+01 # GammaZ [GeV]
    14  0.214100D+01 # GammaW [GeV]
    15  0.116637D-04 # GFermi [GeV]
    16  0.755625D-02 # alphaQED
    17  0.222646D+00 # sin^2(thetaW)
    18  0.508000D-01 # sin^2(thetaC)
Block INTEGER
     1      1 # [1:pp][2:ppbar]
     2      2 # [0:LO][1:NLO][2:NNLO]
     3      1 # [0:WH][1:ZH]
     4      0 # PDF number
     5      5 # nf
     6      1 # [0:w/o][1:with] electro-weak corrections
Block CHAR
     1  MSTW2008nnlo68cl.LHgrid # PDF set
Block SIGMA
     1  0.398313D+00 # sigma(all) [pb]
    10  0.117070D+00 # alpha_s(mz)
    11  0.399203D+00 # sigma(DY) [pb]
    12  0.154259D-01 # sigma(gg->HZ) [pb]
    13  -.510000D+01 # deltaEW [%]
    14  0.404315D-02 # sigma(top) [pb]
\end{verbatim}
}

\newpage
One observes that {\tt vh@nnlo} includes the input to the output file,
and adds a new {\tt Block}, named {\tt SIGMA}. Its entries are given as
follows:
\begin{description}
\item[\tt Block SIGMA] --- {\it output of {\tt vh@nnlo}}\\
\begin{tabular}{rcl}
{\tt 1} &---& \begin{minipage}[t]{.8\textwidth}
{\it $\sigma_\text{tot}$, the total inclusive cross
  section, including all corrections requested in the input file}
\end{minipage}
\\
{\tt 10} &---& 
\begin{minipage}[t]{.8\textwidth}
{\it $\alpha_s(M_\text{Z})$, the input value for the strong
  coupling constant, defined by the \pdf{} set}
\end{minipage}
\\
{\tt 11} &---& 
\begin{minipage}[t]{.8\textwidth}
{\it $\sigma^\text{\dy}$, the \dy{} part of the cross
  section}
\end{minipage}
\\
{\tt 12} &---& 
\begin{minipage}[t]{.8\textwidth}
{\it $\sigma^\gghz{}$, the \gghz{} component of the
  cross section}
\end{minipage}
\\
{\tt 13} &---& {\it $\delta_\text{EW}$, the electro-weak correction factor}\\
{\tt 14} &---& 
\begin{minipage}[t]{.8\textwidth}
    {\it $\sigma^\topdia{}$, the \topdia{} component of the
  cross section}
\end{minipage}
\end{tabular}
\end{description}

Note that {\tt vh@nnlo} will output the results only up to the order
requested in the input file. Thus, for example, $\sigma^\gghz$ and
$\sigma^\topdia{}$ (entries {\tt 12} and {\tt 14}) will be zero unless the
requested order of the calculation is \nnlo{} (i.e.\ {\tt Block
  INTEGER}, entry {\tt 2} is equal to {\tt 2}). 


\subsubsection{Citations}
{\tt vh@nnlo} collects a number of results which need to be acknowledged
in a proper way. It is insufficient for the user of {\tt vh@nnlo} to
simply refer to this manual without explicitly referencing the relevant
literature that entered into the calculation. In order to facilitate
this task, {\tt vh@nnlo} outputs a list of ``texkeys'' in the header of
the output file which have to be
referenced (see Section\,\ref{sec:output}). The full references can be obtained with these texkeys from
the {\sf INSPIRE}~\cite{inspire} database using the {\tt find texkey}
command.


\subsubsection{Parameter scans}

The input file described in section\,\ref{sec:input} defines one single
set of parameters. In order to perform parameter scans, one could modify
the main program {\tt mainfiles/main.f}. However, we recommend leaving
the {\tt Fortran} code of {\tt vh@nnlo} untouched, and rather operate
with scripts such as {\tt slharoutines} which are available from
Ref.\,\cite{slharoutines}.



\section{Conclusions}
{\tt vh@nnlo} is a program that collects the most up-to-date results for
inclusive production of Higgs bosons in association with a weak gauge
boson in the \sm{}. Individual perturbative contributions can be
separately considered. Various parameters can be adjusted. In
particular, parton densities can be easily changed through the use of
the {\tt LHAPDF} library.

We hope that the program will be useful for Higgs physics at the \lhc{}.


\paragraph{Acknowledgments} This work was supported by {\abbrev DFG}, project
HA~2990/5-1, and by the Helmholtz Alliance ``Physics at the Terascale''.


\def\app#1#2#3{{\it Act.~Phys.~Pol.~}\jref{\bf B #1}{#2}{#3}}
\def\apa#1#2#3{{\it Act.~Phys.~Austr.~}\jref{\bf#1}{#2}{#3}}
\def\annphys#1#2#3{{\it Ann.~Phys.~}\jref{\bf #1}{#2}{#3}}
\def\cmp#1#2#3{{\it Comm.~Math.~Phys.~}\jref{\bf #1}{#2}{#3}}
\def\cpc#1#2#3{{\it Comp.~Phys.~Commun.~}\jref{\bf #1}{#2}{#3}}
\def\epjc#1#2#3{{\it Eur.\ Phys.\ J.\ }\jref{\bf C #1}{#2}{#3}}
\def\fortp#1#2#3{{\it Fortschr.~Phys.~}\jref{\bf#1}{#2}{#3}}
\def\ijmpc#1#2#3{{\it Int.~J.~Mod.~Phys.~}\jref{\bf C #1}{#2}{#3}}
\def\ijmpa#1#2#3{{\it Int.~J.~Mod.~Phys.~}\jref{\bf A #1}{#2}{#3}}
\def\jcp#1#2#3{{\it J.~Comp.~Phys.~}\jref{\bf #1}{#2}{#3}}
\def\jetp#1#2#3{{\it JETP~Lett.~}\jref{\bf #1}{#2}{#3}}
\def\jphysg#1#2#3{{\small\it J.~Phys.~G~}\jref{\bf #1}{#2}{#3}}
\def\jhep#1#2#3{{\small\it JHEP~}\jref{\bf #1}{#2}{#3}}
\def\mpl#1#2#3{{\it Mod.~Phys.~Lett.~}\jref{\bf A #1}{#2}{#3}}
\def\nima#1#2#3{{\it Nucl.~Inst.~Meth.~}\jref{\bf A #1}{#2}{#3}}
\def\npb#1#2#3{{\it Nucl.~Phys.~}\jref{\bf B #1}{#2}{#3}}
\def\nca#1#2#3{{\it Nuovo~Cim.~}\jref{\bf #1A}{#2}{#3}}
\def\plb#1#2#3{{\it Phys.~Lett.~}\jref{\bf B #1}{#2}{#3}}
\def\prc#1#2#3{{\it Phys.~Reports }\jref{\bf #1}{#2}{#3}}
\def\prd#1#2#3{{\it Phys.~Rev.~}\jref{\bf D #1}{#2}{#3}}
\def\pR#1#2#3{{\it Phys.~Rev.~}\jref{\bf #1}{#2}{#3}}
\def\prl#1#2#3{{\it Phys.~Rev.~Lett.~}\jref{\bf #1}{#2}{#3}}
\def\pr#1#2#3{{\it Phys.~Reports }\jref{\bf #1}{#2}{#3}}
\def\ptp#1#2#3{{\it Prog.~Theor.~Phys.~}\jref{\bf #1}{#2}{#3}}
\def\ppnp#1#2#3{{\it Prog.~Part.~Nucl.~Phys.~}\jref{\bf #1}{#2}{#3}}
\def\rmp#1#2#3{{\it Rev.~Mod.~Phys.~}\jref{\bf #1}{#2}{#3}}
\def\sovnp#1#2#3{{\it Sov.~J.~Nucl.~Phys.~}\jref{\bf #1}{#2}{#3}}
\def\sovus#1#2#3{{\it Sov.~Phys.~Usp.~}\jref{\bf #1}{#2}{#3}}
\def\tmf#1#2#3{{\it Teor.~Mat.~Fiz.~}\jref{\bf #1}{#2}{#3}}
\def\tmp#1#2#3{{\it Theor.~Math.~Phys.~}\jref{\bf #1}{#2}{#3}}
\def\yadfiz#1#2#3{{\it Yad.~Fiz.~}\jref{\bf #1}{#2}{#3}}
\def\zpc#1#2#3{{\it Z.~Phys.~}\jref{\bf C #1}{#2}{#3}}
\def\ibid#1#2#3{{ibid.~}\jref{\bf #1}{#2}{#3}}
\def\otherjournal#1#2#3#4{{\it #1}\jref{\bf #2}{#3}{#4}}
\newcommand{\jref}[3]{{\bf #1}, #3 (#2)}
\newcommand{\hepph}[1]{{\tt [hep-ph/#1]}}
\newcommand{\mathph}[1]{{\tt [math-ph/#1]}}
\newcommand{\arxiv}[2]{{\tt arXiv:#1}}
\newcommand{\bibentry}[4]{#1, {\it #2}, #3\ifthenelse{\equal{#4}{}}{}{, }#4.}
\bibliographystyle{elsarticle-num}














\end{document}